\documentclass[prc,twocolumn,superscriptaddress,floatfix]{revtex4}
\usepackage{graphics}
\usepackage{epsfig}
\usepackage{amsfonts}
\usepackage{amsmath}
\usepackage{float}
\usepackage{bm}% bold math
\usepackage{mathrsfs}
\usepackage{makecell}
\usepackage{siunitx}
\usepackage{color}
\usepackage{scrextend}
\usepackage{tablefootnote}
\usepackage{footnote}
\usepackage[breaklinks=true,colorlinks=true,linkcolor=blue,urlcolor=blue,citecolor=red]{hyperref}

\begin{document}
	\title{Isoscalar giant monopole strength in $^{58}$Ni, $^{90}$Zr, $^{120}$Sn and $^{208}$Pb 
	}
	\author{A.~Bahini}
	\email[]{a.bahini@ilabs.nrf.ac.za}
	\affiliation{School of Physics, University of the Witwatersrand, Johannesburg 2050, South Africa}
	\affiliation{iThemba Laboratory for Accelerator Based Sciences, Somerset West 7129, South Africa}
	\author{R.~Neveling}
	\email[]{r.neveling@ilabs.nrf.ac.za}
	\affiliation{iThemba Laboratory for Accelerator Based Sciences, Somerset West 7129, South Africa}
	\author{P.~von~Neumann-Cosel}
	\affiliation{Institut f\"{u}r Kernphysik, Technische Universit\"{a}t Darmstadt, D-64289 Darmstadt, Germany}	
	\author{J.~Carter}
	\affiliation{School of Physics, University of the Witwatersrand, Johannesburg 2050, South Africa}
	\author{I.~T.~Usman}
	\affiliation{School of Physics, University of the Witwatersrand, Johannesburg 2050, South Africa}	
	\author{P.~Adsley}
	\altaffiliation[Present address:]{ Department of Physics and Astronomy, Texas A\&M University, College Station, 77843-4242, Texas, USA and Cyclotron Institute, Texas A\&M University, College Station, 77843-3636, Texas USA.}
	\affiliation{School of Physics, University of the Witwatersrand, Johannesburg 2050, South Africa}
	\affiliation{iThemba Laboratory for Accelerator Based Sciences, Somerset West 7129, South Africa}
	\affiliation{\hbox{Department of Physics, Stellenbosch University, Matieland Stellenbosch 7602, South Africa}}
	\affiliation{\hbox{Irene Joliot Curie Lab, UMR8608, IN2P3-CNRS, Universit\'{e} Paris Sud 11, 91406 Orsay, France}}
	\author{N.~Botha}
	\affiliation{School of Physics, University of the Witwatersrand, Johannesburg 2050, South Africa}
	\author{J.~W.~Br\"{u}mmer}
	\affiliation{iThemba Laboratory for Accelerator Based Sciences, Somerset West 7129, South Africa}
	\affiliation{\hbox{Department of Physics, Stellenbosch University, Matieland Stellenbosch 7602, South Africa}}
	\author{L. M. Donaldson}
	\affiliation{iThemba Laboratory for Accelerator Based Sciences, Somerset West 7129, South Africa}
	\author{S.~Jongile}
	\affiliation{iThemba Laboratory for Accelerator Based Sciences, Somerset West 7129, South Africa}
	\affiliation{\hbox{Department of Physics, Stellenbosch University, Matieland Stellenbosch 7602, South Africa}}
	\author{T.~C.~Khumalo}
	\affiliation{School of Physics, University of the Witwatersrand, Johannesburg 2050, South Africa}
	\affiliation{iThemba Laboratory for Accelerator Based Sciences, Somerset West 7129, South Africa}	
	\affiliation{\hbox{Department of Physics, University of Zululand, Richards Bay, 3900, South Africa}}	
	\author{M.~B.~Latif}	
	\affiliation{School of Physics, University of the Witwatersrand, Johannesburg 2050, South Africa}
	\affiliation{iThemba Laboratory for Accelerator Based Sciences, Somerset West 7129, South Africa}
	\author{K.~C.~W.~Li}
	\affiliation{iThemba Laboratory for Accelerator Based Sciences, Somerset West 7129, South Africa}	
	\affiliation{\hbox{Department of Physics, Stellenbosch University, Matieland Stellenbosch 7602, South Africa}}
	\author{P.~Z.~Mabika}
	\affiliation{Department of Physics and Astronomy, University of the Western Cape, Bellville 7535, South Africa}
	\author{P. T. Molema}	
	\affiliation{School of Physics, University of the Witwatersrand, Johannesburg 2050, South Africa}
	\affiliation{iThemba Laboratory for Accelerator Based Sciences, Somerset West 7129, South Africa}	
	\author{C.~S.~Moodley}
	\affiliation{School of Physics, University of the Witwatersrand, Johannesburg 2050, South Africa}
	\affiliation{iThemba Laboratory for Accelerator Based Sciences, Somerset West 7129, South Africa}
	\author{S.~D.~Olorunfunmi}
	\affiliation{School of Physics, University of the Witwatersrand, Johannesburg 2050, South Africa}
	\affiliation{iThemba Laboratory for Accelerator Based Sciences, Somerset West 7129, South Africa}
	\author{P.~Papka}
	\affiliation{iThemba Laboratory for Accelerator Based Sciences, Somerset West 7129, South Africa}
	\affiliation{\hbox{Department of Physics, Stellenbosch University, Matieland Stellenbosch 7602, South Africa}}
	\author{L.~Pellegri}
	\affiliation{School of Physics, University of the Witwatersrand, Johannesburg 2050, South Africa}
	\affiliation{iThemba Laboratory for Accelerator Based Sciences, Somerset West 7129, South Africa}
	\author{B.~Rebeiro}
	\affiliation{Department of Physics and Astronomy, University of the Western Cape, Bellville 7535, South Africa}
	\author{E.~Sideras-Haddad}
	\affiliation{School of Physics, University of the Witwatersrand, Johannesburg 2050, South Africa}
	\author{F.~D.~Smit}
	\affiliation{iThemba Laboratory for Accelerator Based Sciences, Somerset West 7129, South Africa}
	\author{S.~Triambak}
	\affiliation{Department of Physics and Astronomy, University of the Western Cape, Bellville 7535, South Africa}
	\author{M.~Wiedeking}
	\affiliation{School of Physics, University of the Witwatersrand, Johannesburg 2050, South Africa}
	\affiliation{iThemba Laboratory for Accelerator Based Sciences, Somerset West 7129, South Africa}
	\author{J.~J.~van~Zyl}
	\affiliation{\hbox{Department of Physics, Stellenbosch University, Matieland Stellenbosch 7602, South Africa}}
	
	\date{\today}
	\begin{abstract}
		\noindent \textbf{Background:} Inelastic $\alpha$-particle scattering at energies of a few hundred MeV and very-forward scattering angles including $\ang{0}$ has been established as a tool for the study of the isoscalar giant monopole (IS0) strength distributions in nuclei.
		This compressional mode of nuclear excitation can be used to derive the incompressibility of nuclear matter.
		
		\noindent\textbf{Objective:}~An independent investigation of the IS0 strength in nuclei across a wide mass range was performed using the $0^\circ$ facility at iThemba Laboratory for Accelerator Based Sciences (iThemba LABS), South Africa, to understand differences observed between IS0 strength distributions in previous experiments performed at the Texas A\&M University (TAMU) Cyclotron Institute, USA and the Research Center for Nuclear Physics (RCNP), Japan.
		
		\noindent\textbf{Methods:}~The isoscalar giant monopole resonance (ISGMR) was excited in $^{58}$Ni, $^{90}$Zr, $^{120}$Sn and $^{208}$Pb using $\alpha$-particle inelastic scattering with $196$ MeV $\alpha$ beam and scattering angles $\theta_{\text{Lab}} = 0^\circ$ and $4^\circ$.~The K$600$ magnetic spectrometer at iThemba LABS was used to detect and momentum analyze the inelastically scattered $\alpha$ particles.~The IS0 strength distributions in the nuclei studied were deduced with the difference-of-spectra (DoS) technique including a correction factor for the $4^\circ$ data based on the decomposition of $L > 0$ cross sections in previous experiments.
		
		\noindent\textbf{Results:}~IS0 strength distributions for $^{58}$Ni, $^{90}$Zr, $^{120}$Sn and $^{208}$Pb are extracted in the excitation-energy region $E_{\rm x} = 9 - 25$ MeV.~Using correction factors extracted from the RCNP experiments, there is a fair agreement with their published IS0 results.~Good agreement for IS0 strength in $^{58}$Ni is also obtained with correction factors deduced from the TAMU results, while marked differences are found for $^{90}$Zr and $^{208}$Pb.
		
		\noindent\textbf{Conclusions:}~\textcolor{black}{
			Previous measurements show significant differences in the IS0 strength distributions of $^{90}$Zr and $^{208}$Pb. This work demonstrates clear structural differences in the energy region of the main resonance peaks with possible impact on the determination of the nuclear matter incompressibility presently based on the IS0 centroid energies of these two nuclei. The results also suggest that for an improved determination of the incompressibility, theoretical approaches should aim at a description of the full strength distributions rather than the centroid energy only.} 
	\end{abstract}
	
	\maketitle
	
	%=================================================================================================	
	\section{Introduction}
	\label{s1}
	
	The isoscalar giant monopole resonance (ISGMR) is a nuclear collective excitation that can provide information on the bulk properties of the nucleus \cite{Harakeh}.~It was first identified in the late 1970s \cite{harakeh1977mn,youngblood1977isoscalar} and has since then been extensively studied due to its role in constraining the incompressibility of uniform nuclear matter ($K_{\infty}$) \cite{Blaizot,Harakeh,GC_review2018}.~Current knowledge of the ISGMR in stable nuclei depends largely on experimental studies performed at the Texas A\&M University (TAMU) Cyclotron Institute and the Research Center for Nuclear Physics (RCNP) over the past three decades through small-angle (including $\ang{0}$) inelastic $\alpha$-particle scattering measurements at $240$ MeV and $386$ MeV, respectively \cite{GC_review2018}.
	
	There are well-known examples where different systematic trends of the incompressibility of nuclei ($K_A$) are extracted from datasets obtained at these two facilities.~The possibility of nuclear structure contributions to $K_A$ was considered by Youngblood {\it et al.}~\cite{youngblood2013} following the investigation of the ISGMR strength in $^{90,92,94}$Zr and $^{92,96,98,100}$Mo at TAMU.~Such a suggestion would have considerable consequences, since it contradicts the generally held notion that the ISGMR and nuclear incompressibility are collective phenomena and hence, without sensitivity to details of the internal structure of the nucleus.~The ISGMR centroid energy for $^{90}$Zr was reported to be $1.22$ MeV and $2.80$ MeV lower than that for $^{92}$Zr and $^{92}$Mo, respectively, resulting in a value for $K_A$ that increases with mass number.~This unexpected result was subsequently attributed to the high excitation-energy tail of the isoscalar giant monopole (IS0) strengths that were substantially larger in $^{92}$Zr and $^{92}$Mo than for the other Zr and Mo isotopes \cite{krishichayan2015g}.~However, these differences were not observed in independent measurements performed at RCNP.~Using both the difference-of-spectra (DoS) and multipole decomposition analysis (MDA) techniques, it was shown that the ISGMR strengths and energies in $^{90,92}$Zr and $^{92}$Mo are practically identical \cite{gupta2016}.~The study was expanded to include $^{94,96}$Mo \cite{howard2019}, which resulted in the same conclusion based on moment ratios and extracted scaling-model incompressibilities.
	
	Different trends for $K_A$ were also observed for the Ca isotope chain.~Results from ISGMR studies at TAMU for  $^{40,44,48}$Ca \cite{youngblood2001,lui2011,button2017} showed an increase of the ISGMR centroid energy with increasing mass number \cite{button2017}.~In contrast, Howard {\it et al.}~\cite{howard2020} used the experimental facilities at RCNP to study the evolution of the ISGMR strength in $^{40,42,44,48}$Ca and found the generally expected trend of a decrease of the ISGMR centroid energy with increasing mass number.~Recently, Olorunfunmi {\it et al.}~\cite{sunday} presented a third dataset for the Ca isotope chain, obtained at iThemba LABS, and demonstrated that the moment ratios extracted from the three facilities agree when considering an excitation-energy range covering the resonance peak.~It was observed that different trends in the nuclear incompressibility for these nuclei are most likely caused by contributions to the IS0 strength outside of the region covering the resonance peak, and in particular for high excitation energies.
	
	Much of the discussion regarding the source of the different trends in $K_A$ centers around the different background subtraction methods employed by TAMU and RCNP groups \cite{howard2020,gupta2016,GC_review2018} prior to the MDA of the excitation-energy spectra.~The background subtraction methodology used in the TAMU experiments makes assumptions about both the instrumental background and the physical continuum \cite{youngblood2002isoscalar}.~On the other hand, experimental methods employed at RCNP eliminate the instrumental background from the excitation-energy spectra, but contributions from the physical continuum are not distinguished from the IS0 strength in the analysis \cite{GC_review2018}.~In both the Ca and Zr/Mo cases discussed above, comparisons in literature were only made on the basis of trends observed in $K_A$, which is a single number obtained from the ratio of moments of the IS0 strength distribution, that in some cases can be shown to display quite a variation in structural character between different studies.~The existence of such differences led Colo {\it et al.} \cite{colo2020} to conclude that one should rather use the overall shape of the strength distributions in the analysis of the ISGMR instead of the extracted values of the ISGMR energy centroids.~It is, therefore, very important to be aware of the structural variations in the IS0 strength distributions across all available datasets before commenting on the value of, as well as possible trends in $K_A$.
	
	Here, we aim to provide a third measurement of the shape of the IS0 strength distribution in a few medium-to-heavy nuclei in order to extend the comparisons provided in Refs.~\cite{Armand_PRC2022,sunday} for lighter nuclei.	
	\setlength{\parskip}{0pt}
	\section{Experimental details}
	\label{s2}	
	The experimental procedure followed in this study is fully described elsewhere \cite{sunday,Armand_PRC2022}.~As such, only salient details are provided here.~The experiment was performed at the Separated Sector Cyclotron (SSC) facility of the iThemba Laboratory for Accelerator Based Sciences (iThemba LABS) in South Africa.~A beam of $196$-MeV $\alpha$ particles was inelastically scattered off self-supporting $^{58}$Ni, $^{90}$Zr, $^{120}$Sn and $^{208}$Pb targets with areal densities ranging from $0.7$ mg/cm$^2$ to $1.43$ mg/cm$^2$ and isotopically enriched to values greater than $96\%$.~The reaction products were momentum analyzed by the K$600$ magnetic spectrometer \cite{nev11}.~The horizontal and vertical positions of the scattered $\alpha$ particles in the focal plane of the spectrometer were measured using two multiwire drift chambers.~Energy deposition in the plastic scintillators in the focal plane as well as time-of-flight measurements relative to the cyclotron radio frequency were used for particle identification. 
	
\begin{figure*}[t]  %figures 1 and 2
		\begin{minipage}{0.48\textwidth}
			\includegraphics[trim=0.5cm -0.5cm 0 2.5cm, width=\linewidth]{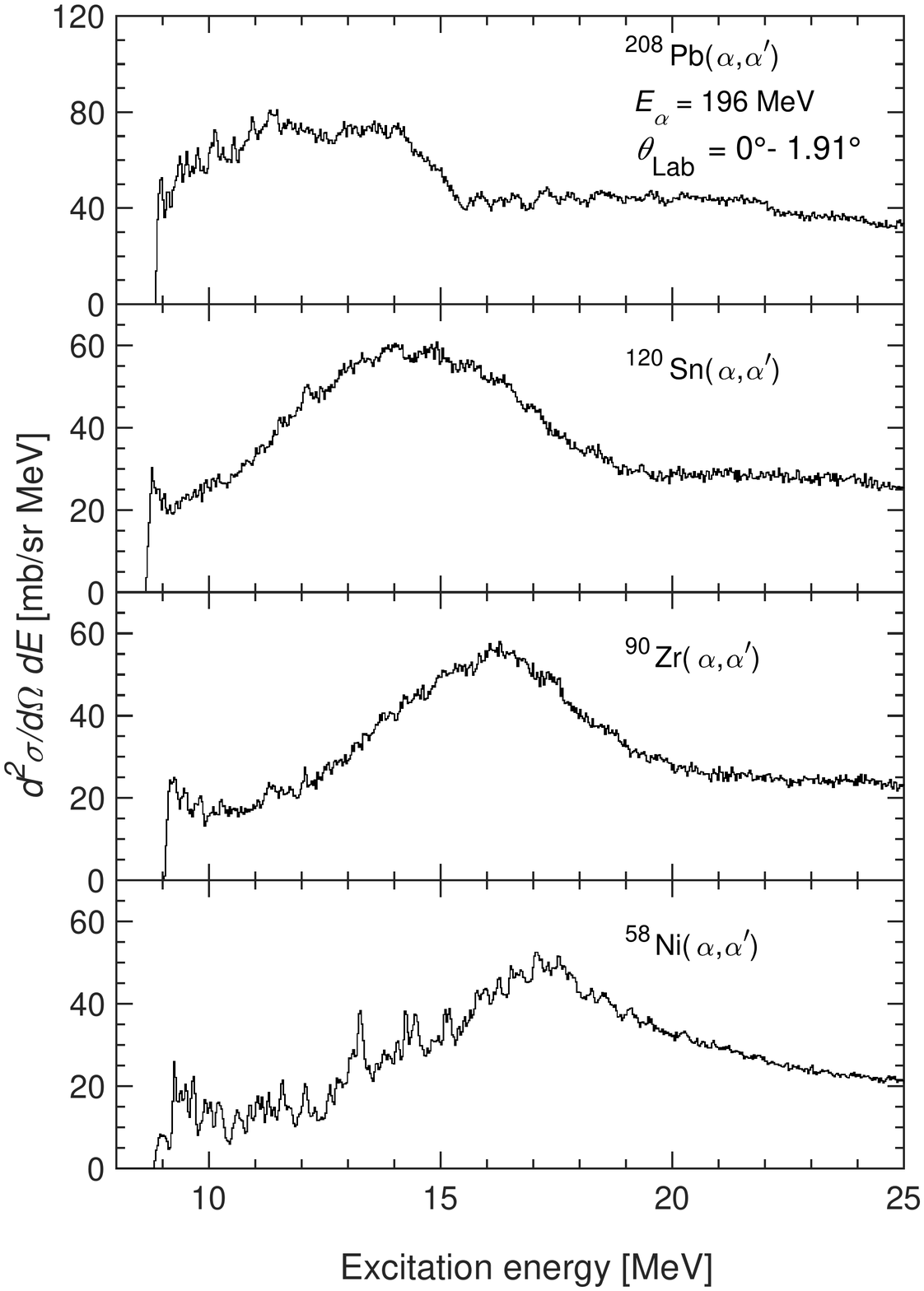}
			\caption{Double-differential cross sections (binned to $30$ keV) for the ($\alpha,\alpha^\prime$) reaction at $E_\alpha = 196$ MeV on $^{208}$Pb,$^{120}$Sn,$^{90}$Zr, and $^{58}$Ni for the angular range $\theta_{\text{Lab}} = 0^{\circ} - 1.91^{\circ}$.}
			\label{FIG:1}
		\end{minipage}\hfill
		\begin{minipage}{0.48\textwidth}
			\includegraphics[trim=0.5cm -0.5cm 0 3.5cm, width=\linewidth]{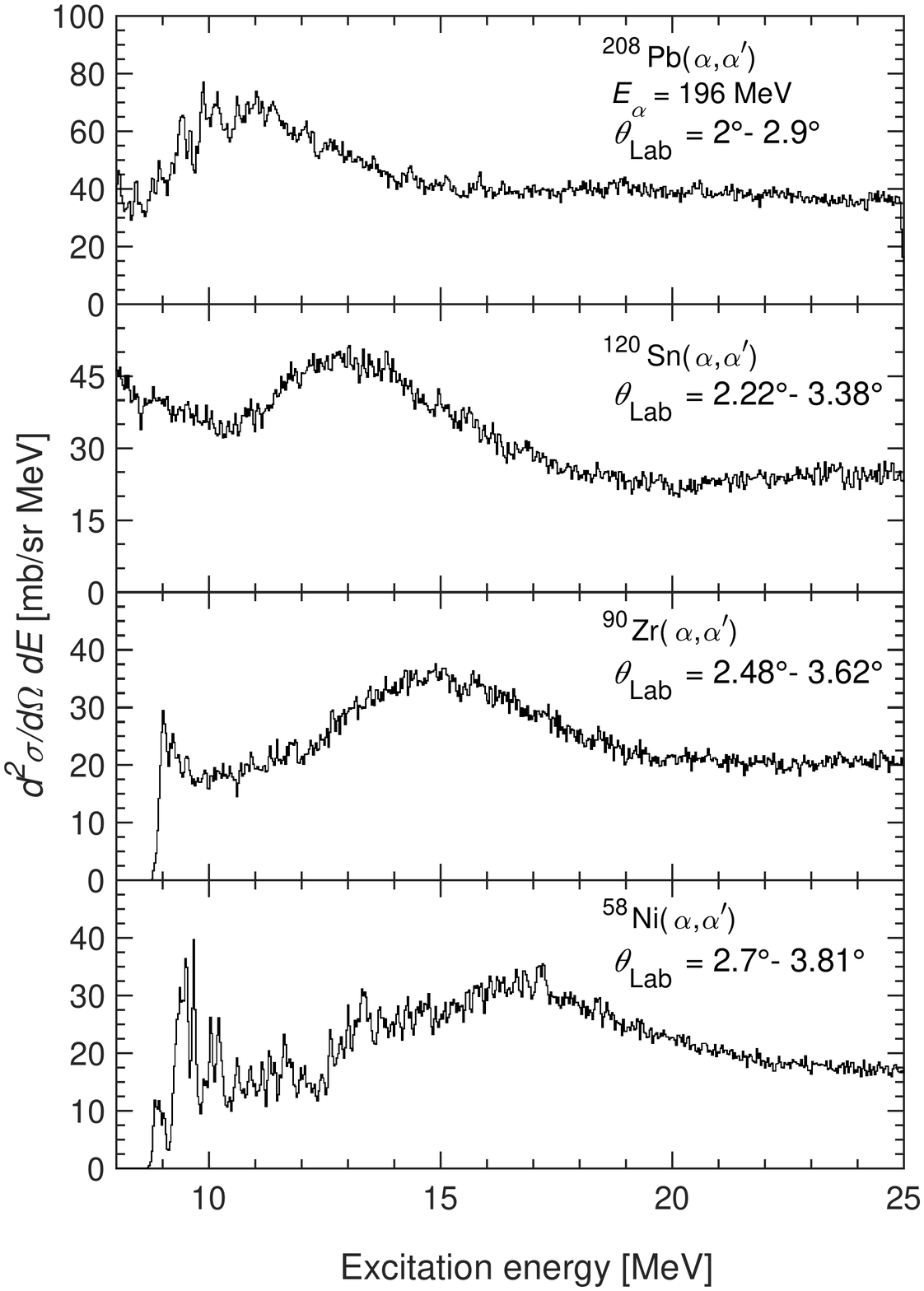}
			\caption{Same as Fig.~\ref{FIG:1}, but for angle cuts as implemented in the $\theta_{\text{K600}} = 4^\circ$ dataset as summarized in Table \ref{table:3}.}
			\label{FIG:2}
		\end{minipage}
	\end{figure*}

	Spectra were acquired with the spectrometer positioned at angles $\theta_{\text{K600}} =0^{\circ}$ and $4^{\circ}$.~In the former, scattering angles of $\theta_{\text{Lab}} = 0^\circ \pm 1.91^{\circ}$ and in the latter, scattering angles from $\theta_{\text{Lab}} = 2^{\circ} - 6^{\circ}$ were covered by a circular spectrometer aperture.~The procedures for particle identification, calibration of the measured focal-plane angles, as well as background subtraction followed those described in Ref.~\cite{Armand_PRC2022}.~The momentum calibration was based on well-known states in $^{24}$Mg \cite{kaw2013,bor1981}, and an energy resolution of $\approx 70$ keV (full width at half maximum, FWHM) was obtained.~Figure \ref{FIG:1} shows the inelastic scattering cross sections extracted 
	at $\theta_{\text{K600}} =0^{\circ}$ for $^{58}$Ni, $^{90}$Zr, $^{120}$Sn and $^{208}$Pb.~Fine structure is clearly observed in the ISGMR region.~The cross sections shown in Fig.~\ref{FIG:2} are for angle ranges that represent a subset of the accessible angle range for the  $\theta_{\text{K600}} = 4^\circ$ measurements, as required to extract monopole strengths.~See Sect.~\ref{s3a} for details.
	
	\textcolor{black}{While the $^{58}$Ni, $^{90}$Zr and $^{120}$Sn target foils were free of contaminants, the $^{208}$Pb target showed signs of surface oxidation.}
	The $12.049$ MeV $J^{\pi}=0^+$ and $11.520$ MeV $J^{\pi}=2^+$ states of $^{16}$O were observed in the $^{208}$Pb spectra measured at $\ang{0}$ and $\ang{4}$, respectively.~While the identifiable peaks of $^{16}$O sit at the lower energy side of the excitation-energy spectrum, $^{16}$O also contributes to the background underneath the ISGMR region.~Therefore, it is essential to remove 
	\textcolor{black}{the contribution from $^{16}$O across the full excitation energy range} prior to the calculation of the differential cross sections.~\textcolor{black}{Accurate $^{16}$O($\alpha,\alpha^\prime$) spectra at $\ang{0}$ and $\ang{4}$ were produced as follows.~Inelastic $\alpha$-particle scattering data from Mylar ($\textrm{C}_{10}\textrm{H}_8\textrm{O}_4$)  and $^{\text{nat}}\textrm{C}$ targets were acquired at $\theta$\textsubscript{lab} = 0$^{\circ}$ and 4$^{\circ}$.~An excitation-energy spectrum for the $^{16}\textrm{O}$($\alpha,\alpha^\prime$) reaction at each angle was then produced by subtracting the $^{12}$C data from the Mylar spectrum, normalized to the 9.641 MeV $J^{\pi}=3^-$ state and the broad \textcolor{black}{resonance strength in this energy region.}~Contributions of the $^{16}\textrm{O}$ contaminant to the excitation-energy spectrum
		of $^{208}$Pb were then removed by subtracting a normalized $^{16}\textrm{O}$ spectrum.
		In the case of the 0$^{\circ}$ dataset the normalization was based on the integrated yield of the $^{16}\textrm{O}$, 12.049 MeV, $J^{\pi}=0^+$ peak and for the 4$^{\circ}$ dataset on the $^{16}\textrm{O}$, 11.520 MeV, $J^{\pi}=2^+$ peak.}
	
	%~An accurate $^{16}$O($\alpha,\alpha^\prime$) spectrum, determined from inelastic $\alpha$-particle scattering data from Mylar ($\textrm{C}_{10}\textrm{H}_8\textrm{O}_4$) and $^{\textrm{nat}}\textrm{C}$ targets measured under the same experimental conditions, was used in order to successfully remove this background from each of the spectra.~\textcolor{black}{In order to achieve this, mylar C$_{10}$H$_8$O$_4(\alpha, \alpha^\prime)$ and carbon $^{12}$C$(\alpha, \alpha^\prime)$ measurements were performed at both $\ang{0}$ and $\ang{4}$.~A normalization of the spectrum from $^{12}$C to the strength of the $^{12}$C $9.641$ MeV peak in the mylar spectrum was performed based on the number of counts in the $9.641$ MeV peak in both mylar and carbon spectra by fitting the peak using a Gaussian plus a polynomial background.~The area under the peak is then obtained which leads to an appropriate scaling.~As such, the $^{12}$C spectrum was removed from the mylar spectrum through direct subtraction to obtain the $^{16}$O($\alpha,\alpha^\prime$) spectrum.}
	
	%=================================================================================================	
	\section{Analysis}
	\label{s3}
	\subsection{DoS technique}
	\label{s3a}
	
	The MDA technique was employed in numerous studies to extract multipole strength distributions in nuclei, including the IS0 strength distributions \cite{GC_review2018,gupta2018isoscalar}.~However, due to the limited number of angular data points in this study, the IS0 strength distributions were determined by means of the DoS technique \cite{DoSpaper}.~This relies on the assumption that the sum of all multipolarity contributions $L >0$ is essentially the same close to $0^\circ$ as at the first minimum of the $L = 0$ angular distribution, and can be removed by subtraction of the spectra measured at the two scattering angles.~The method, therefore, requires the determination of suitable angle cuts  for the different nuclei from the measurement at $\theta_{\text{Lab}} = 2^{\circ} - 6^{\circ}$, which can be assessed from distorted-wave born approximation (DWBA) calculations.
	
	The DoS method requires the prior subtraction of contributions to the spectra due to  relativistic Coulomb excitation of the isovector giant dipole resonance (IVGDR).~The Coulomb cross sections are strongly forward peaked and thus violate the basic DoS assumption.~These contributions are determined using photonuclear cross sections in conjunction with DWBA calculations based on the Goldhaber-Teller model \cite{satchler1987isospin} to estimate the IVGDR differential cross sections as a function of excitation energy.~Lorentzian parameters for the photonuclear cross sections (relative strength $\sigma_m$, peak energy $E_m^{\text{photo}}$ and width $\Gamma^{\text{photo}}$) used in the present study were taken from Ref.~\cite{plujko} and are presented in Table~\ref{table:1}. 
	
	\begin{table}
		\caption{Lorentzian parameters of the photonuclear cross sections from Ref.~\cite{plujko} used for the estimation of Coulomb cross sections at $E_\alpha = 196$ MeV.}	
		\label{table:1}
		\begin{center}
			\setlength{\arrayrulewidth}{0.5pt}
			\setlength{\tabcolsep}{0.15cm}
			\renewcommand{\arraystretch}{1.3}	
			\begin{tabular}{cccc}
				\hline\hline
				Nucleus &  $\sigma_m$ (mb) & $E_m^{\text{photo}}$ (MeV) & $\Gamma^{\text{photo}}$ (MeV)\\
				\hline
				$^{58}$Ni &  $0.294$  & $18.26$ & $6.95$ \\
				$^{90}$Zr &  $0.861$  & $16.84$ & $3.99$ \\
				$^{120}$Sn &  $1.219$  & $15.40$ & $4.86$ \\
				$^{208}$Pb &  $1.121$  & $13.46$ & $3.58$ \\					
				\hline\hline
			\end{tabular}
		\end{center}	
	\end{table} 
	
	\textcolor{black}{In the present study, the DWBA calculations were performed according to the method described in Ref.~\cite{satchler1997missing}.~A density-dependent single-folding model for the real part of the potential $U(r)$, obtained with a Gaussian $\alpha$-nucleon potential, and a phenomenological Woods-Saxon potential for the imaginary term of $U(r)$ were used, so that the $\alpha$-nucleus potential can be written as\\
		\begin{equation}
		\label{e4a}
		U(r) = V_{\text{fold}}(r) + i\dfrac{W}{\left\lbrace 1 + \exp\left[\left(r - R_\text{I}\right)/a_\text{I}\right] \right\rbrace}~,
		\end{equation}\\
		with radius $R_\text{I} = r_\text{0I}(A_{\text{p}}^{1/3}+A_{\text{t}}^{1/3})$ and diffuseness $a_\text{I}$. The subscripts p and t refer to projectile and target, respectively, and $A$ denotes the mass number.~The potential $V_{\text{fold}}(r)$ is obtained by folding the ground-state density with a density-dependent $\alpha$-nucleon interaction\\
		\begin{equation}
		\label{e421b}
		V_{\text{fold}}(r) = - V \int d^3r^\prime \rho(r^\prime)\left[1 - \beta\rho(r^\prime)^{2/3}\right]\exp(- \mathit{z}^2/t^2)~,
		\end{equation}\\
		where $\mathit{z} = |r - r^\prime|$ is the distance between the centre of mass of the $\alpha$ particle and a target nucleon, and $\rho(r^\prime)$ is the ground-state density of the target nucleus at the position $r^\prime$ of the target nucleon.~The parameters $\beta = 1.9$ fm$^2$ and range $t = 1.88$ fm were taken from Ref.~\cite{satchler1997missing}.~The ground-state density $\rho(r)$ of the target nucleus at the position $r$ is given by\\
		\begin{equation}
		\label{e421c}
		\rho(r) = \dfrac{\rho_0}{1 + \exp\left(\frac{r - c}{a}\right)}~,
		\end{equation}\\
		where the Fermi-distribution parameters $c$ and $a$ describe the half-density radius and the diffuseness, respectively.}~Numerical values of the Fermi-distribution parameters $c$ and $a$, which describe the half-density radius and the diffuseness, respectively, were taken from Ref.~\cite{fricke1995nuclear}.~The calculations were carried out using the computer code PTOLEMY \cite{Mac1978,rhoades1980techniques2}.~Optical model parameters used in the DWBA calculations were taken for each nucleus from the studies of the TAMU group on $^{58}$Ni, $^{90}$Zr and $^{116}$Sn.~Here, the $^{116}$Sn nucleus is considered because no result has been published by the group on $^{120}$Sn.~\textcolor{black}{For $^{208}$Pb, elastic scattering cross sections calculated with the parameters quoted in Ref.~\cite{youngblood2004isoscalar} could not reproduce the experimental data of Ref.~\cite{clark2001isoscalar} from which they were said to be derived. Thus, we have performed an independent fit guided by the systematic mass dependence (decrease of real and imaginary depth, increase of imaginary radius) observed for the other nuclei. All parameters are shown in Table \ref{table:2}.
	}
	%For $^{208}$Pb, the single-folded potential used for the real part and the Woods-Saxon potential used for the imaginary part of the optical model were adjusted \textcolor{black}{to improve the fit to} elastic scattering data of Ref.~\cite{clark2001isoscalar}.~The optical model parameters are summarized in Table~\ref{table:2}.
	
	\begin{table}
		\caption{Optical model parameters used in the present study.}	
		\label{table:2}
		\begin{center}
			\setlength{\arrayrulewidth}{0.5pt}
			\setlength{\tabcolsep}{0.15cm}
			\renewcommand{\arraystretch}{1.3}	
			\begin{tabular}{cccccc}
				\hline\hline
				Nucleus &  $V$ (MeV) & $W$ (MeV) & $r_\text{0I}$ (fm) & $a_\text{I}$ (fm) & Refs.\\
				\hline
				$^{58}$Ni &  $41.19$  & $40.39$ & $0.821$ & $0.974$ & \cite{lui2006}\\
				
				$^{90}$Zr &  $40.02$  & $40.9$ & $0.786$ & $1.242
				$ & \cite{krishichayan2015g}\\
				$^{120}$Sn &  $36.7$  & $23.94$ & $0.998$ & $1.047
				
				$ & \cite{youngblood2004isoscalar}\\
				$^{208}$Pb &  $33.3$  & $31.4$ & $1.032$ & $1.057
				
				$ & See text\\
				\hline\hline
			\end{tabular}
		\end{center}	
	\end{table} 
	
	\begin{table}
		\caption{Angle cuts implemented in the $\theta_{\text{K600}} = 4^\circ$ %$\ang{4}$ 
			dataset to define the angular region around the first minimum of the $L = 0$ component ($\theta^{L=0}_{\text{c.m.}}$).}	
		\label{table:3}
		\begin{center}
			\setlength{\arrayrulewidth}{0.5pt}
			\setlength{\tabcolsep}{0.4cm}
			\renewcommand{\arraystretch}{1.3}	
			\begin{tabular}{cccc}
				\hline\hline
				Nucleus & $\theta_{\text{Lab}}$ & $\theta_{\text{c.m.}}$ &  $\theta^{L=0}_{\text{c.m.}}$ \\
				\hline
				$^{58}$Ni & $\ang{2.7}-\ang{3.81}$ & $\ang{2.9}-\ang{4.1}$ & $\ang{3.5}$
				\\
				$^{90}$Zr & $\ang{2.48}-\ang{3.62}$ & $\ang{2.6}-\ang{3.8}$ & $\ang{3.2}$
				\\
				$^{120}$Sn & $\ang{2.22}-\ang{3.38}$ & $\ang{2.3}-\ang{3.5}$ & $\ang{2.9}$
				\\
				$^{208}$Pb & $\ang{2.0}-\ang{2.9}$ & $\ang{2.05}-\ang{2.95}$ & $\ang{2.5}$
				\\
				\hline\hline
			\end{tabular}
		\end{center}	
	\end{table}
	
\begin{figure} %figure 3
		\includegraphics[scale=0.62]{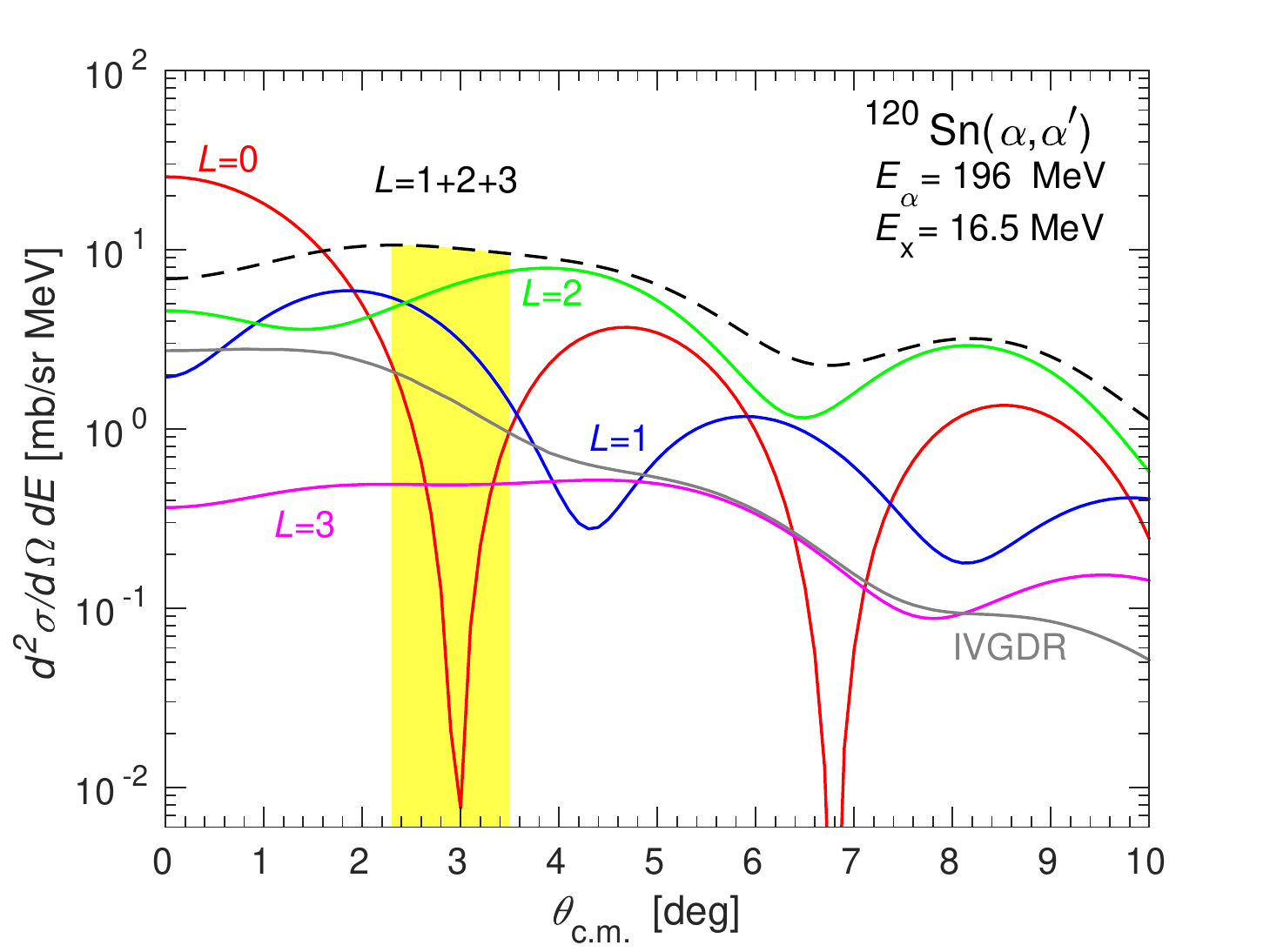}
		\caption{DWBA calculations of the differential cross sections for the $^{120}$Sn($\alpha$,$\alpha^\prime$) reaction at $E_\alpha = 196$ MeV for various isoscalar electric multipoles.~The calculations were done for an excitation energy of $16.5$ MeV, representative of the maximum of the IS0 strength distributions, and scaled using fraction energy-weighted sum rule (FEWSR) strengths from Ref.~\cite{li2010isoscalar}.~The black dashed line represents the sum of all multipoles except $L = 0$.}
		\label{FIG:3}
	\end{figure}

	Consider as an  example the DWBA results for multipoles $L = 0 -3$ as well as the IVGDR cross sections for the case of $^{120}$Sn at an excitation energy of 16.5 MeV, as shown in Fig.~\ref{FIG:3}.~The theoretical angular distributions for excitation of the isoscalar modes are normalized to the corresponding strengths deduced in Ref.~\cite{li2010isoscalar}.~An angular region around the first minimum of the $L = 0$ angular distribution, indicated by the yellow area, is chosen \textcolor{black}{for the subtraction procedure from the zero degree spectrum}.~The angular ranges chosen for the nuclei studied here are summarized in Table~\ref{table:3}.
	
	\begin{figure} %figure 4
		\centering
		\includegraphics[trim=0.05cm -0.5cm 0 3cm, scale=0.47]{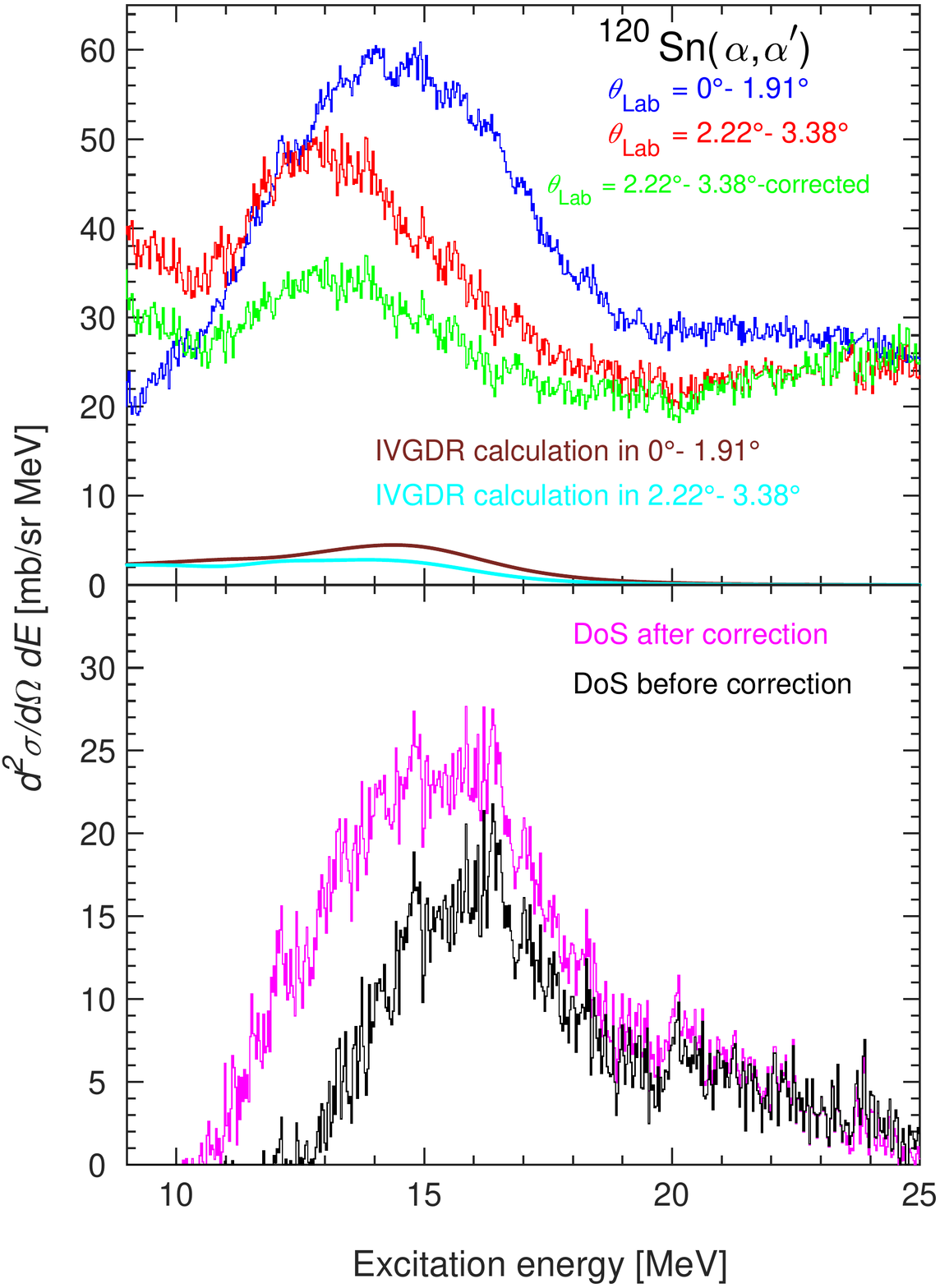}
		\caption{Double-differential cross sections for $^{120}$Sn($\alpha$,$\alpha^\prime$) at $E_\alpha = 196$ MeV.
			Top: The blue and red spectra represent the data acquired at $\ang{0} \leq \theta_{\text{Lab}} \leq \ang{1.91}$ and at $\ang{2.22} \leq \theta_{\text{Lab}} \leq \ang{3.38}$, respectively.~The green spectrum shows the latter spectrum corrected with excitation-energy-dependent factors as outlined in Fig.~\ref{FIG:5} and in the text.~The IVGDR contributions (shown in dark brown and cyan) were subtracted from the blue and red spectra, respectively, prior to the application of the DoS technique.~Bottom: The magenta (black) spectrum represents the difference spectra when applying the DoS technique with (without) the correction factors shown in Fig.~\ref{FIG:5}.
		}
		\label{FIG:4}
	\end{figure}	
	
	A comparison of the spectra extracted from the $0^\circ$ data and the angle cut around the minimum of the $L = 0$ angular distribution for $^{120}$Sn is presented in the upper part of Fig.~\ref{FIG:4} as blue and red histograms, respectively.~The figure also shows the cross sections due to Coulomb excitation of the IVGDR for the two angle settings (brown and cyan lines, respectively).~They are small but non-negligible.~The black histogram in the lower part of Fig.~\ref{FIG:4} is the difference spectrum before the correction procedure described in Sect.~\ref{3b}.

	\subsection{DoS with excitation-energy-dependent corrections}
	\label{3b}
	
	The central premise of the DoS technique is that the sum of the cross sections of all multipoles $L > 0$ 
	\textcolor{black}{is constant at small scattering angles including the region of the first minimum of the $L = 0$ angular distribution \cite{GC_review2018,Armand_PRC2022,DoSpaper}}.~Hence, the subtraction of the inelastic spectrum
	\textcolor{black}{at the angle where $L = 0$ is at a minimum} from the $0^{\circ}$ spectrum is assumed to represent essentially the IS0 component excited in $\alpha$-inelastic scattering close to $0^{\circ}$.~\textcolor{black}{However, as was demonstrated in Ref.~\cite{sunday}, the cross sections from the small-angle measurement can deviate from the sum of all $L > 0$ multipoles in the $\ang{0}$ measurement.}~This is also clear from Fig.~\ref{FIG:3} for the case of $^{120}$Sn.~As such, an excitation-energy-dependent correction factor (CF) to be applied to the small-angle spectrum prior to the application of the DoS technique was introduced, and is written as: \\
	\begin{equation}
	\label{e1}
	\text{CF}(E_\text{x}) = \dfrac{\sum_{L = 1,2,3}\frac{d\sigma^\text{DWBA}}{d\Omega}(E_{\text{x}},\theta^{\text{av.}}_{\text{c.m.}})\Big\vert_L}{\sum_{L = 0,1,2,3}\frac{d\sigma^\text{DWBA}}{d\Omega}(E_{\text{x}},\theta^{L=0}_{\text{c.m.}})\Big\vert_L}~,
	\end{equation}\\
	where $\theta^{\text{av.}}_{\text{c.m.}}$ represents the angle corresponding to the average cross sections between $\theta_{\text{c.m.}} = 0^\circ- 2^{\circ}$, and $\theta^{L=0}_{\text{c.m.}}$ is given in the rightmost column of Table~\ref{table:3}.~The method relies on the availability of information about the relative strengths of the $L > 0$ multipoles from previous measurements on the same nucleus.~Although this makes the procedure model dependent, the results of Ref.~\cite{sunday} indicate that the dependence on the chosen inputs is weak.

	\begin{figure} %figure 5
		\begin{center}
			\includegraphics[trim=0.5cm -0.5cm 0 3.65cm, scale=0.47]{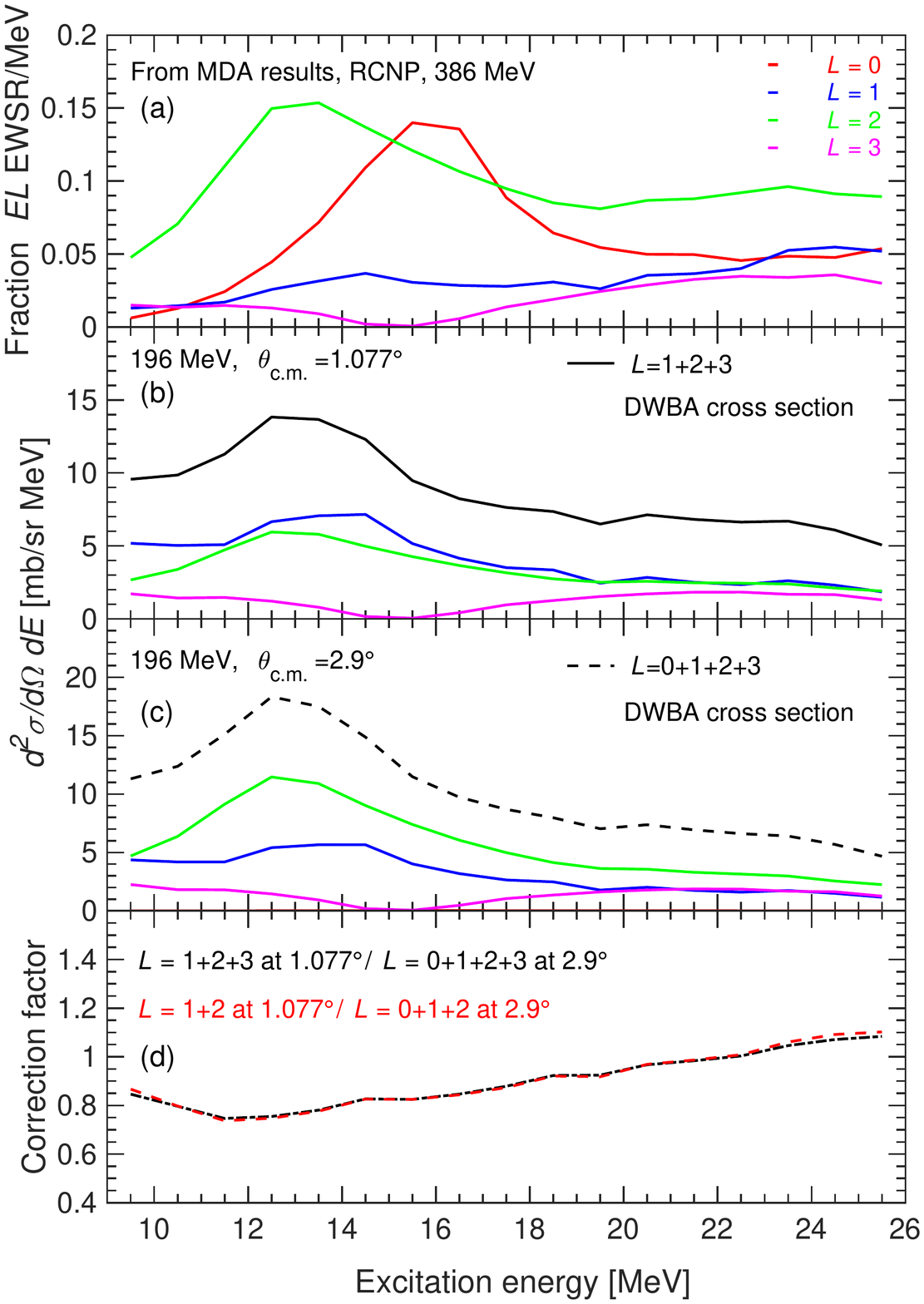}
			
			\caption{Outline of the procedure to establish an excitation-energy-dependent correction factor for the small-angle cross sections, taking $^{120}$Sn as an example.~(a) FEWSR %RN strengths 
				results for different multipoles from RCNP \cite{li2010isoscalar}.~Corresponding DWBA cross sections at $196$ MeV representative of the zero-degree and the small-angle measurements are shown in Panels (b) and (c), respectively.~Panel (d) shows correction factors (black dot-dashed line) determined by the ratio of the $L = 1+2+3$ results in panel (b) to the $L = 0+1+2+3$ results in panel (c).~The red dot-dashed line is the result when $L = 3$ is excluded from the procedure.}
			\label{FIG:5}
		\end{center}
	\end{figure}
	
	The method is illustrated again for the case of $^{120}$Sn in Fig.~\ref{FIG:5}.~Isoscalar $L = 0 - 3$ strength distributions given in  Ref.~\cite{li2010isoscalar} in terms of FEWSR as a function of excitation energy, shown in Fig.~\ref{FIG:5}(a), are used as inputs.~The corresponding DWBA cross sections for $L = 1 - 3$, averaged over the two angular regions of the iThemba LABS experiment, are shown in Figs.~\ref{FIG:5}(b) and (c), respectively.~The summed cross sections are shown as black solid and dashed lines.~We note that Fig.~\ref{FIG:5}(c) also contains an $L = 0$ contribution, as required per Eq.~(\ref{e1}).~However, this contribution is negligibly small (less than $0.1$\% of the summed cross section), which is to be expected as the relevant angle was specifically chosen to be around the minimum of the $L=0$ distribution.~Finally, Fig.~\ref{FIG:5}(d) shows the correction factor as a function of excitation energy, determined from the ratio of the black solid and dashed curves in Figs.~\ref{FIG:5}(b) and (c), respectively.~Application of the correction factors on the small-angle spectrum is shown in the top panel of Fig.~\ref{FIG:4} as a green histogram, and the modified DoS spectrum appears in the bottom panel of Fig.~\ref{FIG:4} as a magenta histogram.~One can see that the correction is particularly strong on the low-energy side of the ISGMR. 
	
\begin{figure} %figure 6
		\begin{center}
			\includegraphics[trim=0.5cm -0.5cm 0 3.6cm, scale=0.47]{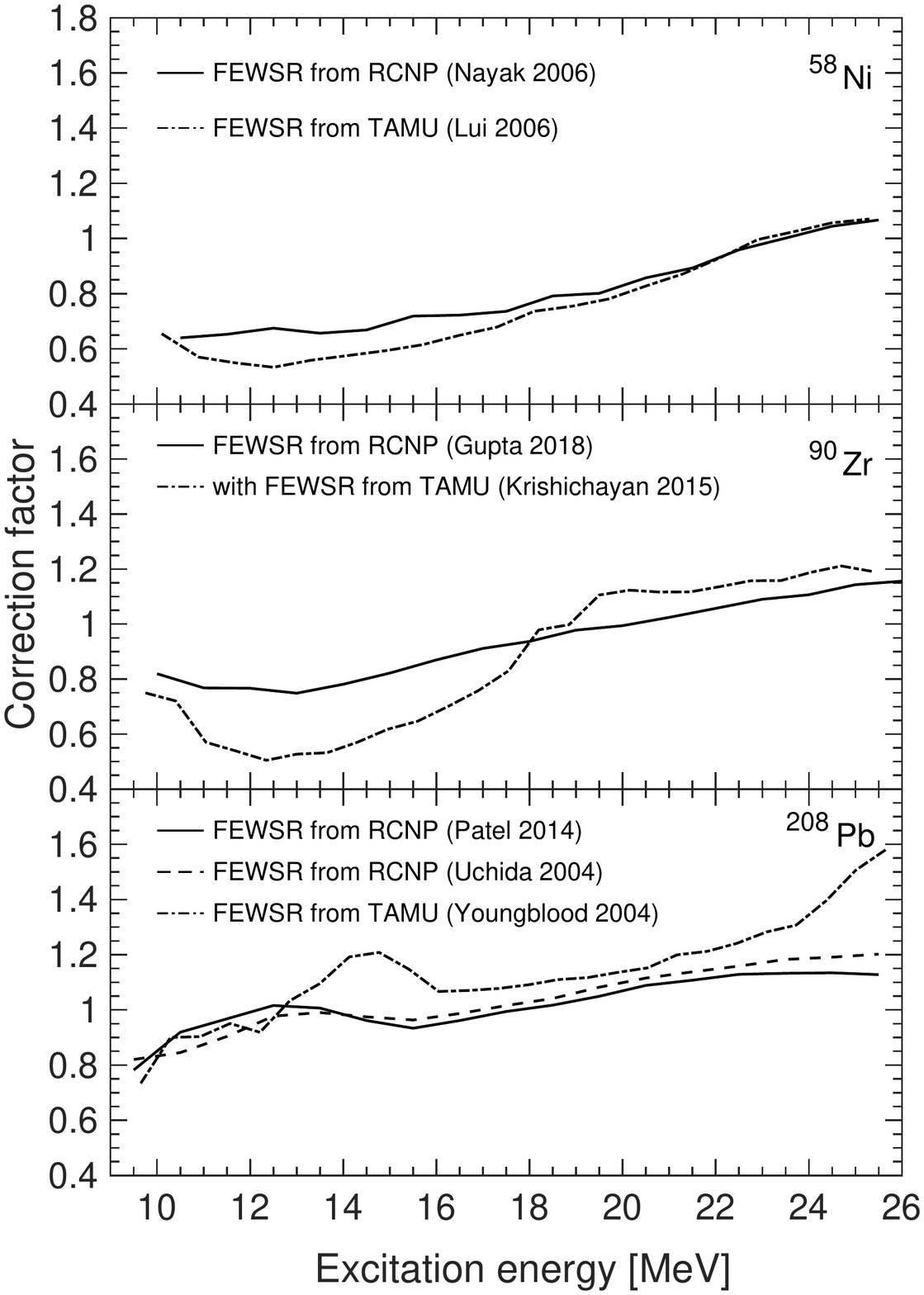}
			
			\caption{Correction factors extracted using FEWSR from RCNP and TAMU datasets, as discussed in the text.~For $^{58}$Ni (top panel), data were taken from \cite{nayak2006} (solid line) and \cite{lui2006} (dashed line); for $^{90}$Zr (middle panel) from \cite{gupta2018isoscalar} (solid line), \cite{howardphd} (dashed line) and \cite{krishichayan2015g} (dash-dotted line); and for $^{208}$Pb (bottom panel) from \cite{patelphd} (solid line), \cite{uchida2004} (dashed line) and \cite{youngblood2004isoscalar} (dash-dotted line).}
			\label{FIG:6}
		\end{center}
	\end{figure}

	The correction factors obtained applying the same procedure to the other nuclei measured in this study are summarized in Fig.~\ref{FIG:6}.~Unlike the case of $^{120}$Sn, where only a single previous measurement was reported, here we have two ($^{58}$Ni and $^{90}$Zr) or even three ($^{208}$Pb) data sets as input, representing both RCNP (solid and dashed lines) and TAMU (dash-dotted lines).~For the case of $^{90}$Zr, correction factors were also determined based on the FEWSR results for the neighboring $^{94}$Mo nucleus \cite{howardphd} assuming that the contributions from different multipolarities change very slowly as a function of nuclear mass number.~Differences between the deduced correction factors are sizable for $^{90}$Zr and $^{208}$Pb when comparing TAMU and RCNP results.~However, the two different experimental results available for both nuclei from RCNP experiments lead to very similar factors.~Thus, only the corrections obtained with Refs.~\cite{howardphd} ($^{90}$Zr) and \cite{patelphd} ($^{208}$Pb) were used to create the RCNP corrected spectra presented and discussed in the next section.
	
	We have investigated to what extent these correction-factor differences may depend on the assumptions in the MDA results of the different experiments, specifically regarding the maximum value of $L$ for which FEWSR results are determined.~This question is discussed in Ref.~\cite{nayak2006} for the case of $^{58}$Ni.~It was found that a variation of $L_{\rm max}$ from $6$ to $8$ had no impact on the FEWSR strengths for $L = 0 - 3$.~A similar conclusion was drawn by Gupta {\it et al.}~\cite{gupta2018isoscalar} for their measurements of $A \approx 90$ nuclei including $^{90}$Zr.~Furthermore, in some of the previously published results, information on the $L = 3$ component is missing.~In general, one expects it to have a minor impact on the correction factors.~The main part of the octupole strength is of a $3 \hbar \omega$ nature and therefore expected at high excitation energies, while its $1 \hbar \omega$ component \cite{fujita1985} is located at excitation energies below the ISGMR.~Nevertheless, we have tested the influence for the case of $^{120}$Sn.~The correction factors obtained by including only $L = 1 + 2$ components are displayed in Fig.~\ref{FIG:4}(d) as a red dashed line.~They fully coincide with the correction factors obtained including $L = 3$ cross sections.~Information on the octupole strength from the investigation of $^{90}$Zr at RCNP \cite{gupta2018isoscalar} is also lacking.~The impact of the $L = 3$ cross sections for the correction factors in this case was estimated from the detailed information on the multipole decomposition analysis provided by Ref.~\cite{howardphd} for the neighboring nucleus $^{94}$Mo and again found to be negligible.

	%=================================================================================================	
	\section{Results and discussion}
	
\begin{figure} %figure 7
		\centering
		\includegraphics[trim=0.3cm -0.5cm 0 3.6cm,width=0.54\textwidth]{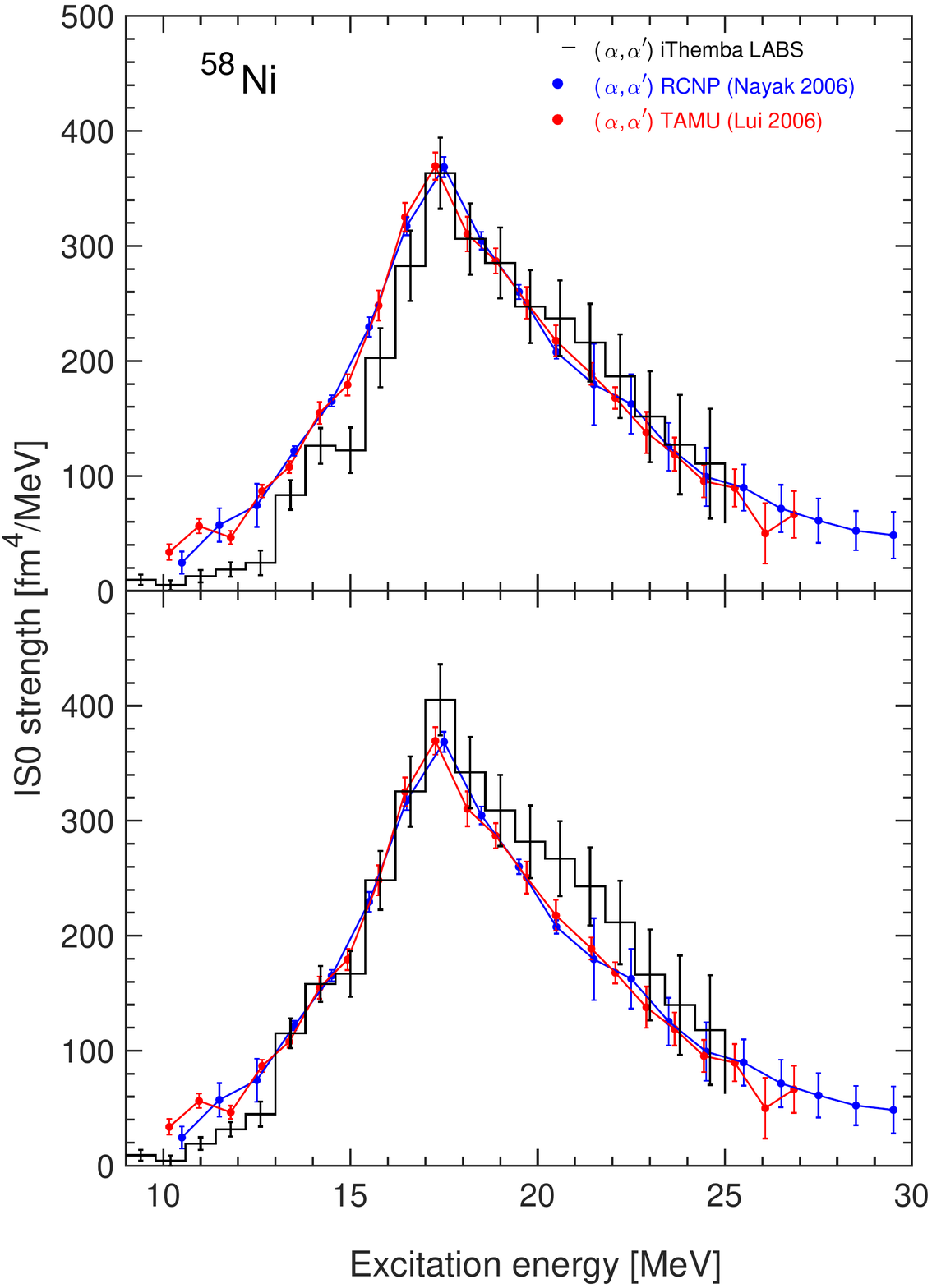}
		\caption{IS0 strength distributions in $^{58}$Ni.~The present iThemba LABS data are shown as black histograms.~Also shown are the ($\alpha, \alpha^\prime$) data from RCNP \cite{nayak2006} (blue filled circles) and TAMU \cite{lui2006} (red filled circles) groups.~The top panel shows results when FEWSR from RCNP are used to correct the small-angle spectrum while the bottom panel displays results when FEWSR from TAMU are used to correct the small-angle spectrum.}
		\label{FIG:7}
	\end{figure}

\begin{figure} %figure 8
		\centering
		\includegraphics[trim=0.3cm -0.5cm 0 3cm,width=0.54\textwidth]{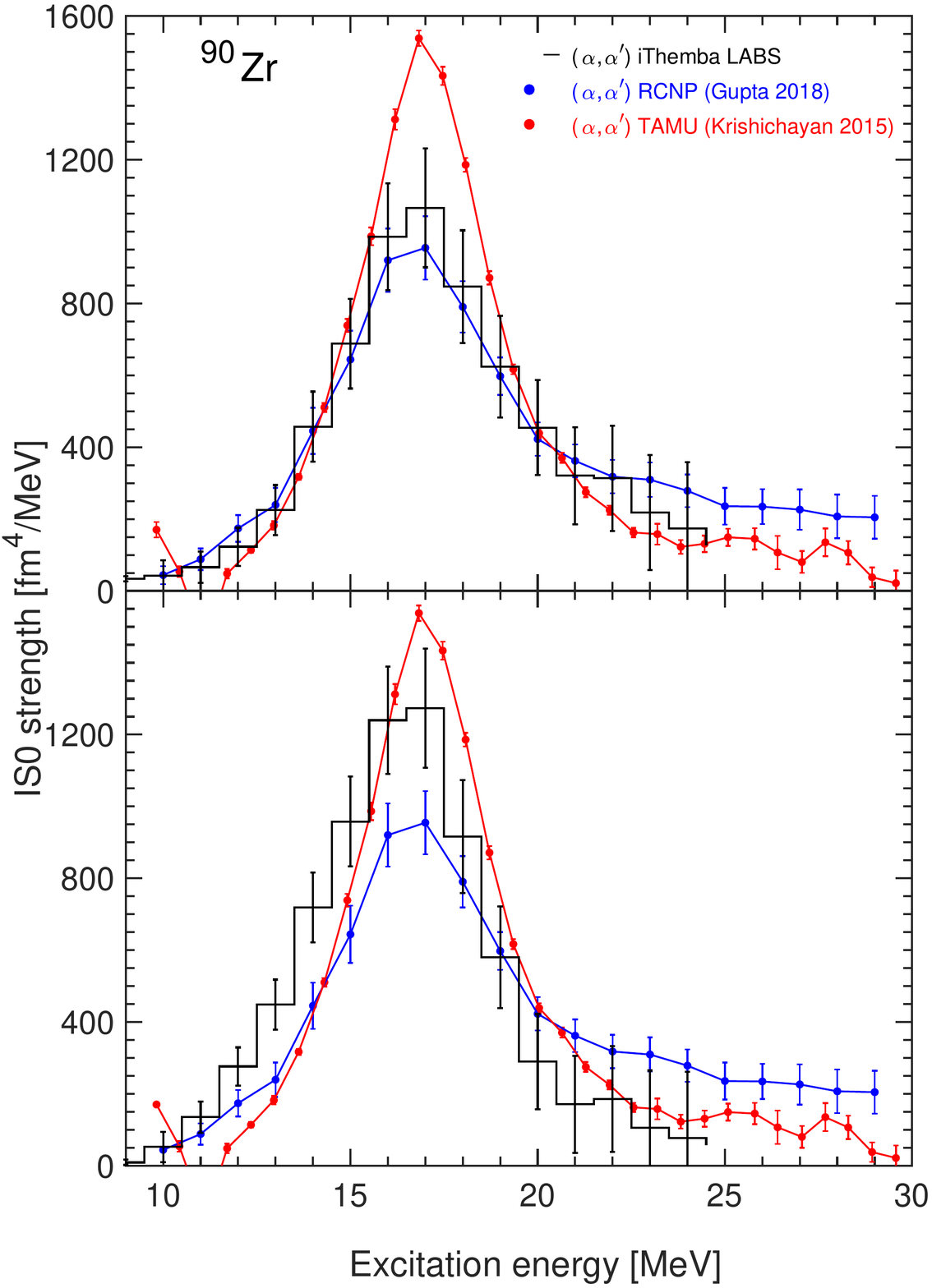}
		\caption{Same as Fig.~\ref{FIG:7} but for $^{90}$Zr.~Also shown are the ($\alpha, \alpha^\prime$) data from RCNP \cite{gupta2018isoscalar} (blue filled circles) and TAMU \cite{krishichayan2015g} (red filled circles).}
		\label{FIG:8}
	\end{figure}

\begin{figure} %figure 9
		\centering
		\includegraphics[trim=0.27cm -0.5cm 0 0.1cm, width=0.5\textwidth]{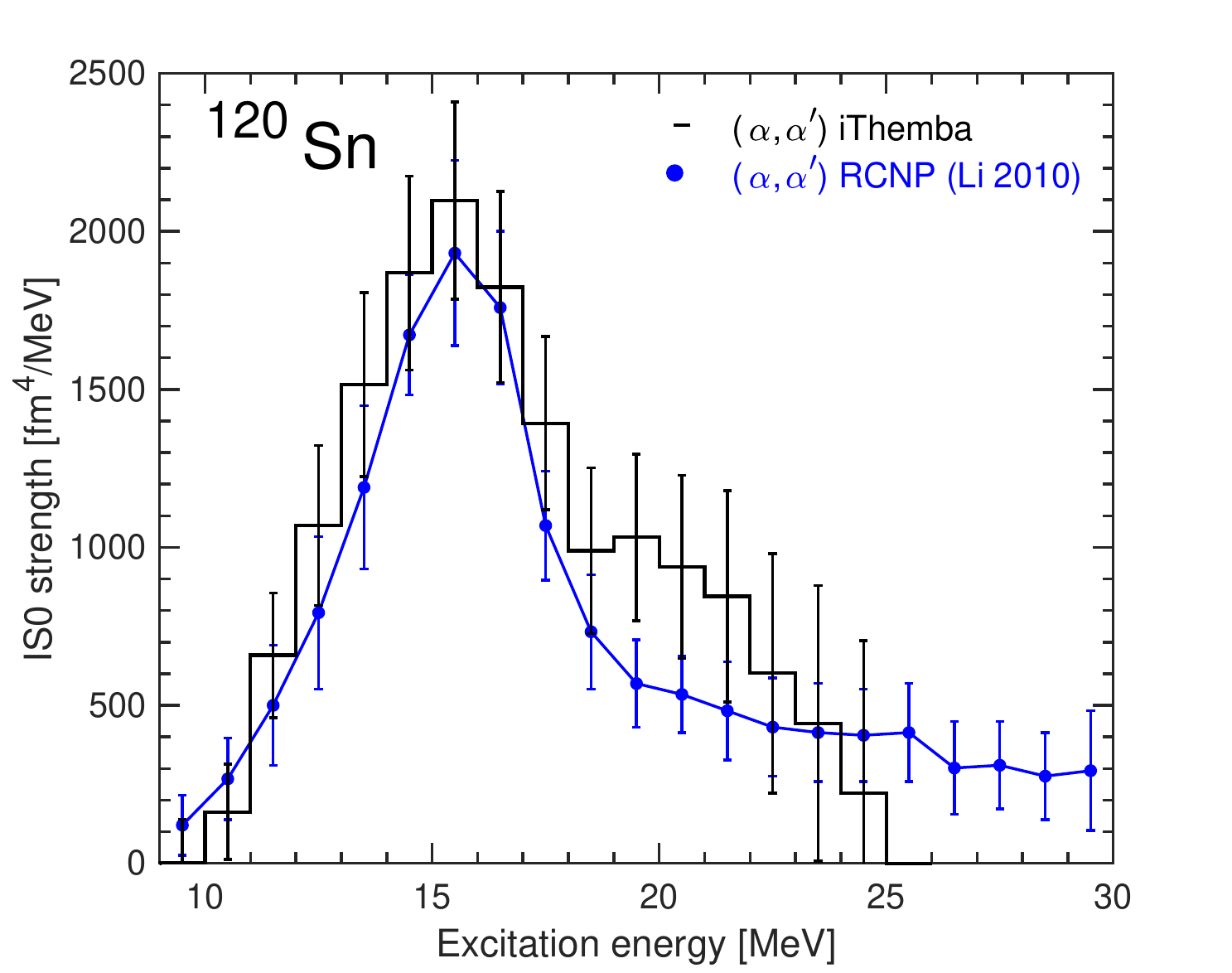}
		\caption{Same as Fig.~\ref{FIG:7} but for $^{120}$Sn.~Also shown are the ($\alpha, \alpha^\prime$) data from RCNP \cite{li2010isoscalar} (blue filled circles).~Here, only FEWSR from RCNP are used to correct the small-angle spectrum.}
		\label{FIG:9}
	\end{figure}

\begin{figure} %figure 10
		\centering
		\includegraphics[trim=0.4cm -0.5cm 0 3.4cm,width=0.54\textwidth]{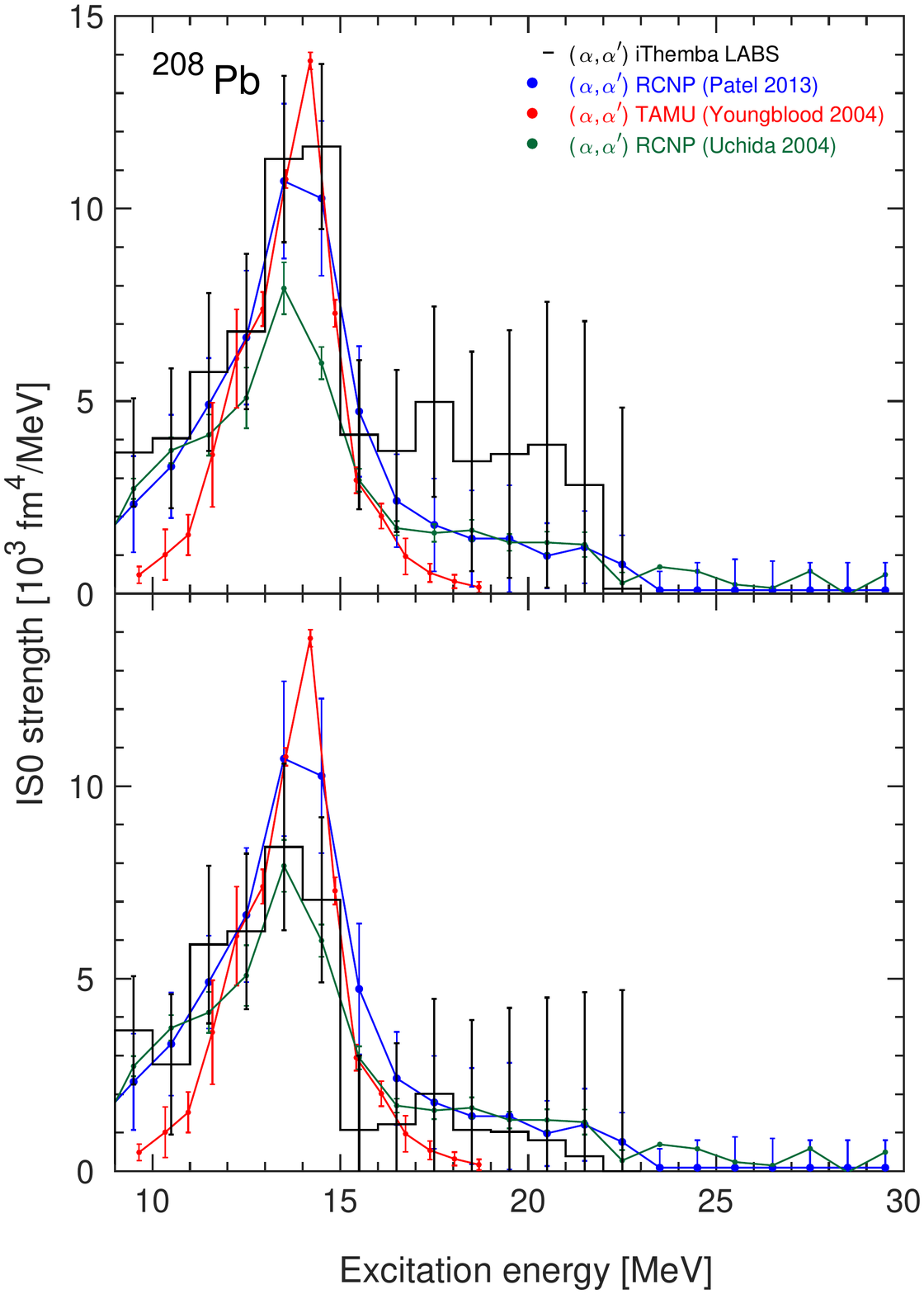}
		\caption{~Same as Fig.~\ref{FIG:7} but for $^{208}$Pb.~Also shown are the ($\alpha, \alpha^\prime$) data from RCNP \cite{uchida2004,patel2013testing} (blue and dark green filled circles) and TAMU \cite{youngblood2004isoscalar} (red filled circles) groups.}
		\label{FIG:10}
	\end{figure}

	The corrected difference cross sections can be converted to fractions $a_0(E_{\text{x}})$ of the isoscalar monopole EWSR by comparing with DWBA calculations assuming $100\%$ EWSR, as shown in Ref.~\cite{sunday}.~The IS0 strength was determined using a $1$ MeV bin size for all nuclei except $^{58}$Ni, which  was binned to $800$ keV, to facilitate direct comparison with previous experiments, as shown in Figs.~\ref{FIG:7}-\ref{FIG:10}, using the following equation \cite{GC_review2018}:\\
	\begin{equation}
	\label{e5}
	S_{0}(E_\text{x}) = \frac{\text{EWSR(IS0)}}{E_\text{x}} a_0(E_\text{x})=\dfrac{2\hbar^2 A \langle r^2\rangle}{mE_\text{x}}a_0(E_\text{x})~.
	\end{equation}\\
	Here, $m$ represents the nucleon mass, $E_\text{x}$ is the excitation energy, and $\langle r^2\rangle$ is the second moment of the ground-state density.~The values for $\langle r^2\rangle$ for $^{58}$Ni, $^{90}$Zr, $^{120}$Sn, and $^{208}$Pb were derived from Ref.~\cite{fricke1995nuclear} and found to be  $14.3$, $18.2$, $21.7$, and $30.3$ fm$^2$, respectively.~Note that all results from TAMU were originally presented as fractions $a_0(E_{\text{x}})$, and were therefore converted to IS0 strength following the same procedure.
	
	The IS0 strength distributions for $^{58}$Ni are presented in Fig.~\ref{FIG:7}, where the iThemba LABS results shown in the upper and lower panels were extracted using correction factors derived from RCNP \cite{nayak2006} and TAMU \cite{lui2006} experiments, respectively.~Here, as is the case for the other results from iThemba LABS, the errors associated with the strength distributions include both systematic and statistical uncertainties.~The IS0 strength distributions from the two previous experiments agree within error bars.~The iThemba LABS strength distribution is in reasonable agreement with the two previous datasets, regardless of the choice of correction factor.~Slightly weaker strengths are seen in the lower excitation-energy region $9$ $ \leq E_\text{x} \leq 16$ MeV when utilizing the RCNP-based correction factor.~On the other hand, using the TAMU-based correction factor, the distribution is somewhat stronger than the TAMU and RCNP distributions in the high  excitation-energy region $20 \leq E_\text{x} \leq 25$ MeV.
	
	The case for $^{90}$Zr is summarized in Fig.~\ref{FIG:8}.~The original controversy in the mass $90$ region was attributed to the high excitation-energy tail of the IS0 strength that is substantially larger in $^{92}$Zr and $^{92}$Mo in the TAMU experiment than for the other Zr and Mo isotopes \cite{krishichayan2015g}.~However, here we clearly see that there are also significant structural differences at the peak of the resonance between data from RCNP \cite{gupta2018isoscalar} and TAMU \cite{krishichayan2015g}.~The IS0 strength from the present experiment utilizing the RCNP correction factors is in very good agreement with the results from RCNP.~On the other hand, when the correction factors are based on the results from TAMU, the centroid of the IS0 distribution shifts to a lower excitation energy.~While the absolute value of the strength at its peak undergoes a noteworthy increase, the overall agreement with previous datasets deteriorates.
	
	The comparison of the IS0 strength distribution in $^{120}$Sn from the RCNP experiment \cite{li2010isoscalar} with the present analysis is presented in Fig.~\ref{FIG:9}.~There is good agreement for the main part of the ISGMR up to about $18$ MeV and at higher excitation energies.~Between $19$ and $22$ MeV, the present results indicate a larger strength, just outside the $1\sigma$ error bars.
	
	\begin{table*}[t]
		\caption{Parameters extracted from the ISGMR strength distributions from previous ($\alpha,\alpha^\prime$) measurements through Lorentzian or Gaussian peak fitting as well as moment ratio calculations, established over different excitation-energy ranges.}	
		\label{table:4}
		\begin{center}
			\setlength{\arrayrulewidth}{0.5pt}
			\setlength{\tabcolsep}{0.3cm}
			\renewcommand{\arraystretch}{2}	
			\begin{tabular}{cccccccc}
				\hline\hline
				Nucleus &  \makecell{Centroid \\ (MeV)} & \makecell{Width \\ (MeV)} & \makecell{$m_1/m_0$ \\ (MeV)} & \makecell{$\sqrt{m_1/m_{-1}}$ \\ (MeV)} & \makecell{$\sqrt{m_3/m_{1}}$ \\ (MeV)} & \makecell{Energy range \\ (MeV)} & Reference  \\
				\hline		
				%\hline
				\vspace{0.2cm}%\hline 
				$^{58}$Ni &  \makecell{\rule{0.5cm}{0.02 cm} \\ $18.43 \pm 0.15$} & \makecell{ \rule{0.5cm}{0.02 cm} \\ $7.41 \pm 0.13$} & \makecell{$19.9^{+0.7}_{-0.8}$ \\ $19.20^{+0.44}_{-0.19}$} & \makecell{\rule{0.5cm}{0.02 cm} \\ $18.70^{+0.34}_{-0.17}$} &\makecell{\rule{0.5cm}{0.02 cm} \\ $20.81^{+0.90}_{-0.28}$} & \makecell{$10.5 - 32.5$ \\ $10 - 35$} & \makecell{RCNP \cite{nayak2006} \\  TAMU \cite{lui2006}\footnote{Peak positions and widths (FWHM) from Gaussian fits.}} \\
				\vspace{0.2cm}%\hline
				$^{90}$Zr & \makecell{$16.76 \pm 0.12$ \\ $17.1  $} & \makecell{$4.96^{+0.31}_{-0.32}$ \\ $4.4$} & \makecell{$19.17^{+0.21}_{-0.20}$ \\ $17.88^{+0.13}_{-0.11}$} &\makecell{$18.65 \pm 0.17$ \\ $17.58^{+0.06}_{-0.04}$} & \makecell{$20.87^{+0.34}_{-0.33}$ \\ $18.86^{+0.23}_{-0.14}$} & \makecell{$10 - 30$ \\ $10 - 35$} &  \makecell{RCNP \cite{gupta2018isoscalar}\footnote{Peak positions and widths (FWHM) from Lorentzian fits.} \\ TAMU\cite{krishichayan2015g}$^{a}$ } \\
				\vspace{0.2cm}%\hline
				$^{120}$Sn &  \makecell{$15.4 \pm 0.2$} & \makecell{$4.9 \pm 0.5$} & \makecell{$15.7 \pm 0.1$} & \makecell{$15.5 \pm 0.1$} & \makecell{$16.2 \pm 0.2$}& \makecell{$10.5 - 20.5$} & \makecell{RCNP \cite{li2010isoscalar}$^{b}$} \\
				\vspace{0.2cm}
				$^{208}$Pb &  \makecell{$13.7 \pm 0.1$ \\ $13.4 \pm 0.2$ \\ \rule{0.5cm}{0.02 cm}} & \makecell{$3.3 \pm 0.2$ \\ $4.0 \pm 0.4$ \\ $2.88 \pm 0.20$\footnote{The equivalent Gaussian FWHM.}} 
				& \makecell{\rule{0.5cm}{0.02 cm}  \\ \rule{0.5cm}{0.02 cm}  \\ $13.96 \pm 0.20$} & \makecell{$13.5 \pm 0.1$ \\ \rule{0.5cm}{0.02 cm} \\ \rule{0.5cm}{0.02 cm}} & \makecell{\rule{0.5cm}{0.02 cm}  \\ \rule{0.5cm}{0.02 cm}  \\ \rule{0.5cm}{0.02 cm}}& \makecell{$9.5 - 19.5$ \\ $8 - 33$ \\ $10 - 35$}  & \makecell{RCNP \cite{patel2013testing}$^{b}$ \\ RCNP \cite{uchida2004}$^{b}$ \\TAMU \cite{youngblood2004isoscalar}} \\
				\hline\hline
			\end{tabular}
		\end{center}	
	\end{table*} 
	
	\begin{table*}[t]
		\caption{Lorentzian parameters and moment ratios for the ISGMR strength distributions in $^{58}$Ni, $^{90}$Zr, $^{120}$Sn, and $^{208}$Pb, where $m_k = \int E_{\text{x}}^k S(E_{\text{x}}) dE_{\text{x}}$ is the $k$th moment of the strength distribution for the excitation-energy range $10 - 24.5$ MeV ($10 - 17$ MeV  for $^{208}$Pb) from the present work, compared to values extracted for the TAMU and RCNP data sets over the same excitation-energy range. 
		}	
		\label{table:5}
		\begin{center}
			\setlength{\arrayrulewidth}{0.5pt}
			\setlength{\tabcolsep}{0.4cm}
			\renewcommand{\arraystretch}{2}	
			\begin{tabular}{ccccccc}
				\hline\hline
				Nucleus &  \makecell{Centroid \\ (MeV)} & \makecell{Width \\ (MeV)} &  \makecell{$m_1/m_0$ \\ (MeV)} & \makecell{$\sqrt{m_1/m_{-1}}$ \\ (MeV)} & \makecell{$\sqrt{m_3/m_{1}}$ \\ (MeV)} & Reference  \\
				\hline	
				%\hline
				\vspace{0.2cm}%\hline 
				$^{58}$Ni &  \makecell{$17.8 \pm 0.4$ \\ $17.8 \pm 0.4$ \\ $17.9 \pm 0.3$ \\ $17.9 \pm 0.3$} & \makecell{$5.4 \pm 0.4$ \\ $5.4 \pm 0.4$ \\ $5.3 \pm 0.3$ \\ $5.3 \pm 0.3$} & \makecell{$18.40 \pm 0.15$ \\ $18.22 \pm 0.13$ \\ $18.15 \pm 0.11$ \\ $18.14 \pm 0.06$} & \makecell{$18.14 \pm 0.14$ \\ $17.94 \pm 0.13$ \\ $17.85 \pm 0.11$ \\ $17.81 \pm 0.06$} & \makecell{$19.12 \pm 0.17$ \\ $18.98 \pm 0.15$ \\ $19.00 \pm 0.12$ \\ $19.00 \pm 0.06$} & \makecell{Present, CF from \cite{nayak2006} \\ Present, CF from \cite{lui2006} \\ RCNP \cite{nayak2006} \\  TAMU \cite{lui2006}} \\
				\vspace{0.2cm}%\hline
				$^{90}$Zr & \makecell{$16.7 \pm 0.2$ \\ $16.2 \pm 0.2$ \\ $16.8 \pm 0.2$ \\ $16.9 \pm 0.2$} & \makecell{$4.4 \pm 0.2$ \\$4.2 \pm 0.2$ \\ $4.8 \pm 0.3$ \\ $3.9 \pm 0.3$} & \makecell{$17.06 \pm 0.35$ \\ $16.02 \pm 0.36$ \\ $17.59 \pm 0.11$ \\ $17.23 \pm 0.03$} & \makecell{$16.80 \pm 0.32$ \\ $15.79 \pm 0.32$ \\ $17.31 \pm 0.11$ \\ $17.03 \pm 0.03$} & \makecell{$17.84 \pm 0.48$ \\ $16.69 \pm 0.57$ \\ $18.41 \pm 0.11$ \\ $17.81 \pm 0.04$} & \makecell{Present, CF from \cite{howardphd} \\ Present, CF from \cite{krishichayan2015g} \\ RCNP \cite{gupta2018isoscalar} \\ TAMU \cite{krishichayan2015g}} \\
				\vspace{0.2cm}%\hline
				$^{120}$Sn &  \makecell{$15.5 \pm 0.4$ \\ $15.4 \pm 0.2$} & \makecell{$5.6 \pm 0.4$ \\$4.6 \pm 0.3$} & \makecell{$16.24 \pm 0.39$ \\ $16.54 \pm 0.23$} & \makecell{$15.92 \pm 0.35$ \\ $16.20 \pm 0.22$} & \makecell{$17.21 \pm 0.54$ \\ $17.61 \pm 0.25$} &\makecell{Present, CF from \cite{li2010isoscalar} \\ RCNP \cite{li2010isoscalar}} \\
				\vspace{0.2cm}
				$^{208}$Pb &  \makecell{$13.8 \pm 0.3$ \\ $13.3 \pm 0.3$ \\  $13.7 \pm 0.2$ \\ $13.4 \pm 0.2$ \\ $13.9 \pm 0.3$} & \makecell{$3.1 \pm 0.2$ \\$3.2 \pm 0.3$ \\ $3.4 \pm 0.2$ \\ $4.0 \pm 0.3$ \\$2.3 \pm 0.4$} & \makecell{$13.39 \pm 0.27$ \\ $12.44 \pm 0.45$ \\  $13.47 \pm 0.22$ \\ $13.78 \pm 0.29$ \\ $13.64 \pm 0.08$} & \makecell{$13.25 \pm 0.26$ \\ $12.29 \pm 0.42$ \\  $13.32 \pm 0.22$ \\ $13.59 \pm 0.27$ \\ $13.56 \pm 0.08$} & \makecell{$13.80 \pm 0.29$ \\ $12.90 \pm 0.57$ \\  $13.86 \pm 0.22$ \\ $14.32 \pm 0.35$ \\ $13.85 \pm 0.07$} &\makecell{Present, CF from \cite{patel2013testing} \\ Present, CF from \cite{youngblood2004isoscalar} \\ RCNP \cite{patel2013testing} \\ RCNP \cite{uchida2004} \\TAMU \cite{youngblood2004isoscalar}} \\
				\hline\hline
			\end{tabular}
		\end{center}	
	\end{table*} 
	
	For the case of $^{208}$Pb, results from three different previous experiments \cite{patel2013testing,uchida2004,youngblood2004isoscalar} are available.~The IS0 strength distributions from these studies are compared with one another in Fig.~\ref{FIG:10}.~Upon inspection of the different strength distributions, it is clear that there are distinct structural differences between the different datasets.~Youngblood {\it et al.}~\cite{youngblood2004isoscalar} produced a very narrow IS0 distribution that is not nearly as asymmetric as the results from both Uchida~{\it et al.}~\cite{uchida2004} and Patel {\it et al.}~\cite{patel2013testing}.~The TAMU study also reported the highest value for the monopole strength at the peak of the distribution, while the strength at the peak is almost a factor of two lower in Ref.~\cite{uchida2004}, while the results from Ref.~\cite{patel2013testing} lie in between.~The latter two distributions reach their maximum at a slightly lower excitation energy.~The iThemba LABS results corrected using the FEWSR results available from Ref.~\cite{patel2013testing} are in fair agreement with the IS0 distribution from that paper.~On the other hand, it better agrees with the IS0 distribution from Ref.~\cite{uchida2004} when the TAMU-based correction factors are employed.~The strength visible above $17$ MeV in the present data (and eventually also in the high excitation region of the $^{120}$Sn data) might be attributed to a less than perfect subtraction of the low-energy flank of the ISGDR \cite{uchida2003} that dominates the background cross sections.
	
	A total of $83 \pm 5 \%$ $(96 \pm 5 \%)$, $84 \pm 9 \%$ $(88 \pm 9 \%)$, $112 \pm 11 \%$, and $124 \pm 14 \%$ $(85 \pm 14 \%)$ of the IS0 EWSR was identified for $^{58}$Ni, $^{90}$Zr, $^{120}$Sn, and $^{208}$Pb using the RCNP- (TAMU)-based correction factors.~The quoted EWSR fractions have been calculated over the excitation-energy range $10 - 24.5$ MeV \textcolor{black}{($10 - 17$ MeV for $^{208}$Pb)}, encompassing the main ISGMR peak and the errors associated include both systematic and statistical uncertainties.~While a comparison to previously quoted values is difficult because they strongly depend on the chosen energy interval, they illustrate that most of the ISGMR strength is found in the energy range covered by the present data.
	
	There are clear structural differences between results originating from TAMU and RCNP in the case of $^{90}$Zr and $^{208}$Pb, but not for $^{58}$Ni. These differences are within the main region of the ISGMR, and not confined to high excitation energies where background subtraction effects might be expected to dominate.~In the case of $^{58}$Ni, the iThemba LABS data show fair agreement with previous datasets regardless of the source of the correction factors, but the picture is unfortunately not so clear for the heavier nuclei.~This is due to the reliance in this study on the $L>0$ strength distributions sourced from the very experiments with which we wish to compare IS0 results.~Consider that for $^{90}$Zr  there is, at best, agreement between iThemba LABS and RCNP results when using the RCNP-based correction factor and, at worst, a situation of three distinct IS0 strength distributions.~For the case of $^{208}$Pb, the iThemba LABS data either agrees with the strength distribution from Uchida~{\it et al.}~\cite{uchida2004} or Patel {\it et al.}~\cite{patel2013testing} depending on the use of the TAMU- or RCNP-based correction factors, respectively.
	
	It is interesting to consider the various values used to characterize the energy of the ISGMR reported in the literature for the data shown in Figs.~\ref{FIG:7}-\ref{FIG:10}, originating either from peak fitting or from moment ratio calculations \cite{Lipp89}.~The results are summarized in Table \ref{table:4}.~Clearly, the value assigned to the ISGMR centroid depends on the calculation method.~Peak fitting with Gaussian or Lorentzian distributions is not very satisfactory, as the real shape of the ISGMR rarely conforms to these simplistic peak shapes.~Values for the various moment ratios, on the other hand, depend heavily on the excitation-energy range over which they are calculated, and in the absence of clear guidelines, one finds quite a variation in the integration ranges utilized.~The behavior of the scaling model energies ($\sqrt{m_3/m_{1}}$) for the case of $^{58}$Ni and $^{90}$Zr as compared to $^{120}$Sn confirms the impact of large integration ranges on the extracted centroid values.~The higher values of the RCNP results \cite{gupta2018isoscalar} in the case of $^{90}$Zr, even for a smaller excitation-energy range covered than in the TAMU results \cite{krishichayan2015g}, stem from possible contributions due to the physical continuum at high excitation energies \cite{GC_review2018}.
	
	It is important to be aware of these complications, as differences of several hundred keV impact on the extraction of nuclear matter incompressibility from theoretical calculations.~For example, in Ref.~\cite{Shl2006} the value for $K_{\infty}$ was constrained by using the $m_1/m_0$ ratio of the TAMU data to represent the energies of the ISGMR in $^{208}$Pb and $^{90}$Zr.~The experimental centroid energies typically change by 
	%$200 - 300$ keV if the $\sqrt{m_1/m_{-1}}$ ratio was taken instead. 
	\textcolor{black}{more than 400 keV if the results from RCNP studies are used instead, as done in 
		a recent study by Li {\it et al.} \cite{Li2022}, where the centroid for $^{208}$Pb originates from the moment ratio $\sqrt{m_1/m_{-1}}$. }
	
	While the structural differences in the strength distributions highlighted in previous paragraphs will also contribute towards the range of values reported in Table \ref{table:4}, the large variations in applicable energy ranges make it impossible to compare these results on an even footing.~For this reason we calculated, for all the strength distributions shown in Figs.~\ref{FIG:7}-\ref{FIG:10}, the three moment ratios over the same excitation-energy range, and present the results in Table \ref{table:5}.~In addition, we fitted the IS0 strength distributions with a Lorentzian\\
	\begin{equation}
	\label{e6}    
	S\left(E_\text{x}\right) = \dfrac{\sigma_0}{\left(E_\text{x}^2 - E^2_0\right)^2 + E_\text{x}^2\Gamma^2}~,
	\end{equation}\\
	in order to extract characteristic centroid and width parameters.~Here, $E_0$ and $\Gamma$ represent the peak energy and width of the resonance, and $\sigma_0$ denotes the strength value at $E_0$.~These show large variations from the moment ratios, demonstrating again that the ISGMR strength distributions are not well approximated by a Lorentzian shape. 
	\textcolor{black}{
		The results in Table V confirm that differences up to several hundred keV in centroid energies calculated through any of the moment ratio methods can be observed between the available datasets.
	}
	%\textcolor{black}{The results in Table \ref{table:4} confirms that differences of several hundred keV in centroid energies calculated through any of the moment ratio methods are observed between the different datasets, clearly indicating how the structural differences can have an effect on the extracted values of the centroid energy.}
	
	%\textcolor{black}{It is therefore clear that comparisons between different experimental studies as well as theory should not be based on a single number, i.e. the centroid energy that is subject more often than not to different integration ranges or calculation methods, but also on the structure of the strength distribution.} 
	\textcolor{black} {It is, therefore, clear that the comparison between different experimental studies as well as theory should not be based only on a single number, i.e., the centroid energy dependent on energy  integration ranges or calculation methods, but also on the full strength distributions. This view is supported by recent studies \cite{GC_review2018,colo2020}.
		%Recent theoretical studies have paid attention to these problems and indicated that one should look at the full strength distribution rather than the energy centroid only to quantify whether calculations reproduce the experimental findings \cite{GC_review2018,colo2020}.
		Theoretically, this} requires going beyond the mean-field level in calculations and include at least particle-vibration coupling (PVC).~A current study of the ISGMR in the chain of stable tin isotopes including PVC \cite{Li2022} demonstrates centroid shifts of several hundred keV potentially resolving the longstanding problem that the random-phase approximation (RPA) calculations require a significantly lower value of $K_\infty$ to describe the Sn isotopes than $^{208}$Pb.

	%=================================================================================================	
	\section{Conclusions}
	\label{s4}	
	We present IS0 strength distributions on nuclei over a wide mass range obtained with the DoS method modified to allow for excitation-energy-dependent correction factors.~These were deduced from available information on $L > 0$ isoscalar strengths.~The need for input from other experiments introduces a model dependence in the analysis.~When using input from various previous studies the effects were found to be negligible for $^{58}$Ni, but large for $^{90}$Zr and $^{208}$Pb.~In general, when taking the $L >0$ strengths from RCNP experiments, fair to good agreement with the IS0 strength distributions from those experiments is achieved.
	
	There is quite a variation in values of ISGMR centroids reported in literature.~Besides the much discussed problems of the subtraction of an empirical background (containing physical and instrumental parts) favored by the TAMU group and the possible inclusion of $L = 0$ strength unrelated to the ISGMR at high excitation energies in the analysis of RCNP data, we show that the structural differences in the main ISGMR peak in results from previous experiments impact on the centroid energy.~This is particularly true in the cases of $^{90}$Zr and $^{208}$Pb, which have been used to extract the nuclear matter incompressibility from the comparison to RPA calculations with different forces.  
	
	While the present data cannot resolve the experimental issues, because of the model-dependent method of extraction of the IS0 strength, they underline a need for new high-precision data on key nuclei for the determination of $K_\infty$ combined with an improved theoretical treatment aiming at a description of the full strength distributions rather than the ISGMR centroids only.~Theoretically, this requires the inclusion of complex configurations beyond the level of RPA.~As an example, a current study of the ISGMR in the chain of stable tin isotopes demonstrates centroid shifts of several hundred keV when PVC is included \cite{Li2022}, allowing for a consistent description with forces reproducing the centroid in $^{208}$Pb. 
	
	\setlength{\parskip}{0pt}	
	%\begin{acknowledgements}
	\section*{ACKNOWLEDGEMENTS}
	
	The authors thank the Accelerator Group at iThemba LABS for the high-quality dispersion-matched beam provided for this experiment.~We are indebted to  G.~Col\`{o} and U.~Garg for useful discussions.~This work was supported by the Deutsche Forschungsgemeinschaft under contract SFB $1245$ (Project ID No.~$79384907$) and by an NRF-JINR grant JINR200401510986.~A.B. acknowledges financial support through iThemba LABS, NRF South Africa.~R.N. acknowledges support from the NRF through Grant No.~$85509$.~P.A. acknowledges support from the Claude Leon Foundation in the form of a postdoctoral fellowship.~This work is based on the research supported in part by the National Research Foundation of South Africa (Grant Number: $118846$).
	
	%\end{acknowledgements}

	%=================================================================================================	

\end{document}